\documentclass{aamas2017}

\usepackage{times}
\usepackage{array, multirow}
\usepackage{graphicx}
\usepackage{latexsym}
\usepackage{amssymb}
\usepackage{fancyvrb}
\usepackage{amsmath} 
\usepackage{mathtools}
\usepackage{centernot}
\usepackage{algpseudocode}
\usepackage{algorithm}
\usepackage{xcolor}
\usepackage{soul}
\usepackage{tikz}
\usetikzlibrary{trees}
\usepackage{caption,subcaption}
\usepackage{bm}
\usepackage{url}
\usepackage{paralist}
\usepackage{hyperref}

\algnotext{EndFor}
\algnotext{EndIf}
\algnotext{EndFunction}
\algnotext{EndProcedure}

\DeclareCaptionType{copyrightbox}

\newtheorem{mydef}{Def.}
\newtheorem{myProbl}{Problem}
\newtheorem{mylemma}{Lemma}
\newtheorem{myproof}{Proof.}
\newtheorem{example}{Example}

\begin{document}

\title{Automating decision making to help establish norm-based regulations}

\numberofauthors{3}

\author{
\alignauthor
\hspace{-0.5in}
Maite Lopez-Sanchez\\
\hspace{-0.5in}
Marc Serramia\\
 \hspace{-0.5in}	
\affaddr{Math \& Comp Science dept.}\\
 \hspace{-0.5in}	
\affaddr{Universitat de Barcelona}\\
 \hspace{-0.5in}	
\affaddr{maite\_lopez@ub.edu}\\
 \alignauthor
\hspace{-0.3in}
Juan A. Rodriguez-Aguilar\\
\hspace{-0.3in}  
\affaddr{Artificial Intelligence}\\
 \hspace{-0.3in}      
\affaddr{Research Institute (IIIA-CSIC)}\\
\hspace{-0.3in}      
\affaddr{Campus UAB. Bellaterra, Spain}\\
\hspace{-0.3in}      
\affaddr{jar@iiia.csic.es}\\
 \alignauthor
\hspace{-0.4in}		
Javier Morales and\\
\hspace{-0.4in}
Michael Wooldridge\\
 \hspace{-0.4in}   
\affaddr{Dept. of Computer Science}\\
\hspace{-0.4in}    
\affaddr{University of Oxford, UK}\\
\hspace{-0.4in}
\affaddr{javier.morales@cs.ox.ac.uk}\\
\hspace{-0.4in}    
\affaddr{mjw@cs.ox.ac.uk}
 %
}

\maketitle

\begin{abstract}
Norms have been extensively proposed as coordination mechanisms for both agent and human societies. Nevertheless, choosing the norms to regulate a society is by no means straightforward. The reasons are twofold. First, the norms to choose from may not be independent (i.e, they can be related to each other). Second, different preference criteria may be applied when choosing the norms to enact. This paper advances the state of the art by modeling a series of decision-making problems that regulation authorities confront when choosing the policies to establish. 
In order to do so, we first identify three different norm relationships --namely, generalisation, exclusivity, and substitutability-- and we then consider norm representation power, cost, and associated moral values as alternative preference criteria.
Thereafter, we show that the decision-making problems faced by policy makers can be encoded 
as linear programs, and hence solved with the aid of state-of-the-art solvers.
\end{abstract}

\vspace{-0.06in}

\ccsdesc[500]{Computing methodologies~Multi-agent systems}

\ccsdesc[500]{Computing methodologies~Cooperation and coordination}

\printccsdesc

\keywords{Normative systems, value-based reasoning, norm decision making, policy making, optimisation.}

\section{Introduction}
\label{sec:intro}

Norms have been extensively studied as coordination mechanisms within both agent and human societies\cite{BoellaTV06, sethi1996evolution}. Within agent societies, problems such as norm synthesis \cite{ShohamT95, agotnes10aamas}, norm emergence \cite{griffiths10coin,VillatoroSS11}, or norm learning \cite{savarimuthu2013prohibitions, Campos13, Riveret:2014} have been widely studied. As for human societies, e-participation and e-governance ICT systems are currently attracting a lot of attention \cite{weerakkody:2012:public,DeTar:2013PhD,Loomio:2016}. Thus, for example, some regulatory authorities in European cities --such as Reykjavik\cite{Rey:2016:RPP} or Barcelona\cite{BCN:2016:BPP} municipalities-- are opening their policy making to citizens. This is also the case for some countries: New Zealand authorities are opening consultations about legislations related to different topics such as family violence\cite{NewZealand:2016:FV} or pensions\cite{NewZealand:2016:P}. However, the number of regulations to discuss and enact could be large --consider, for example, Madrid's participation portal \cite{Madrid:2016}, which is currently hosting more than two thousand local proposals open to citizens-- so that managing them becomes a complex task.        

Beyond the intrinsic complexity due to the number of norms to manage --either if they are proposed by humans or automatically generated--, by no means we can state that choosing the norms to regulate a society constitutes a straightforward process. The reasons are twofold. 

On the one hand, norms can be related. Norm relationships have been previously studied in the literature. Thus, for example, Grossi and Dignum \cite{DignumAbsNorms} study the relation between abstract and concrete norms, whereas Kollingbaum, Vasconcelos et al. \cite{Kollingbaum2007,VasconcelosKN09} focus on norm conflicts ---and solve them based on first-order unification and constraint solving techniques.
In this paper we borrow some of the relationships identified in Morales et al. \cite{morales2015liberal}\footnote{Morales et al.  \cite{morales2015liberal} identify substitutability, generalisation and complementarity relations.} and characterise three different binary norm relationships, namely, generalisation, exclusivity, and substitutability. Thus, we can consider a set of norms and the fact that some norms in this set generalise some specific norms; that some other norms are pair-wise incompatible (i.e., mutually exclusive); or interchangeable (that is, substitutable). 
When this is the case, a regulatory authority  should not select these norms to be simultaneously established in the society. This paper proposes to encode these relationships in terms of restrictions in linear programs that allows to find those norm subsets (subsets of the given set of norms) that are compliant with the constraints imposed by the associated norm relations.

On the other hand, this paper also characterises the problems that regulation authorities confront when considering different preference criteria over the norms to impose. In this manner, we specify the optimisation problem of finding the subset of norms that, in addition to comply with the relation constraints, maximizes represented norms. This problem can be specified as a single objective function in a binary linear program. Moreover, since  norms have associated costs, it may also be of convenience to specify a multi-objective decision function that maximizes norm representation while minimizing associated norm costs. Our final contribution is the consideration of  moral values associated to norms as an additional criterion. Values have been studied in argumentation --some representative examples being Bench-Capon et al.\cite{Bench-CaponA09} or Modgil \cite{Modgil:2006}-- and they have also been introduced by Kohler et al. \cite{Kohler:2014} in multi-agent institutions. However, to the best of our knowledge, no one has considered the values that norms support. Instead, the normative multi-agent systems research area has focused in different normative concepts such as minimality and simplicity \cite{fitoussi00choosing, MoralesAAMAS2014}, liberality \cite{morales2015liberal}, compactness \cite{morales2015compact}, or stability \cite{sethi1996evolution}.

In this paper we assume that a regulatory authority has available a collection of norms to impose together with an specification of the particular relationships that hold for these norms and that prevents all norms to be simultaneously deployed. Then, we model alternative problems that pursue to maximize the set of norms to establish under (a combination of) those previously mentioned criteria, namely, norm representation, associated costs, and supported moral values. \textcolor{black}{Subsequently, despite the computation complexity of these problems, state-of-the-art linear programming solvers are used to automatically compute their solution.}

The paper is structured as follows. First, next section provides some basic definitions and an illustrative example. Then, subsequent sections characterise our different optimisation problems. 
Initially, Section \ref{sec:maxNormSystem} considers optimisation problems with a single objective: maximising norm representation. Then, Section \ref{sec:OptProbl} considers a multi-objective decision function that combines norm representation and cost criteria. Finally, Section \ref{sec:ValProbl} introduces moral values into this multi-objective optimisation problem. The paper concludes with additional discussion about dealing with norms in force in Section \ref{sec:discussion}, and conclusions and future work in Section \ref{sec:concl}. 

\section{Basic definitions}
\label{sec:def}


Before introducing the decision problems involving norms that we face in this paper, next we introduce the fundamental building blocks to build such problems. Thus, we formally introduce the notion of norm, the relationships that we consider between norms, and the characteristic of the norm systems that we will aim at to establish norm-based regulations.

Our notion of norm is based on a simplification of the one in \cite{LopezLI02}. Thus, here we formally consider a norm $n_i$ as a pair $\theta(\rho,ac)$, where:  $\theta$ is a deontic operator (prohibition, permission, or obligation) ; $\rho$ is a description of the addressee entity, namely, the agent required to comply with the norm; and $ac$ is an action --from a set of actions-- that entities can perform in a specific domain.


Now we consider the relations that may hold between norms because they will determine the way norms are selected as a part of a norm system. With this aim, we borrow two of the relations empirically identified in \cite{morales2015liberal} during on-line norm synthesis and define three norm relations (namely \emph{generalisation}, \emph{exclusivity}, and \emph{substitutability}). Informally, it is considered that: (i) a norm is more general than another one when it subsumes its regulation (its regulation scope is wider); (ii) two norms are mutually exclusive when they are incompatible; and (iii) two norms are substitutable if they are interchangeable. 

Let $N$ be a non-empty set of norms for a specific domain. Formally, we will capture the above norm relations as follows:
%
%
\begin{enumerate}[i)]
\item The \textit{direct generalisation} relation is a binary relation $R_g \subseteq N \times N$. If $(n_i,n_j) \in R_g$, we say that $n_i$ is more general than $n_j$, or directly, it generalises $n_j$. Notice also that if $(n_i,n_j) \in R_g$, $\nexists n_k \in N$ s.t $(n_i,n_k),(n_k,n_j)  \in R_g$. The notion of direct generalisation allows us to capture the notion of indirect generalisation through the so-called \emph{ancestors}. Given two norms, $n_k$,$n_i \in N$, we say that $n_k$ is an ancestor of $n_i$ if there is a  subset of norms $\{n_1, \ldots, n_p\} \subseteq N$ such that 
$(n_1,n_2), \ldots, (n_{p-1},n_p) \in R_g$, $n_1=n_k$, and $n_p=n_i$. Henceforth, given a norm $n_i \in N$, we will note its ancestors as $\mathcal{A}(n_i)$.
Notice that $R_g$ is irreflexive, anti-symmetric, and intransitive. 

\item The \textit{exclusivity} relation is a binary relation $R_x \subseteq N \times N$. If $(n_i,n_j) \in R_x$ we say that $n_i, n_j$ are incompatible or mutually exclusive. 
$R_x$ is an irreflexive, symmetric, and intransitive relation.
 
\item The \textit{substitutability} relation is a binary relation $R_s \subseteq N \times N$. If $(n_i,n_j) \in R_s$, we say that norms $n_i, n_j$ are interchangeable or substitutable. Based on substitutability relationships,  
we introduce the notion of \emph{substitution chain} as follows. Given two norms, $n_i$,$n_k \in N$, we say that $n_k$ is \textit{connected by substitutabilities} to $n_i$ if there is a non-empty subset of norms $\{n_1, \ldots, n_p\} \subseteq N$ such that $(n_1,n_2), \ldots,$ $(n_{p-1},n_p$)$\in R_s$, $n_1=n_i$, and $n_p=n_k$. Henceforth,
a new relationship $\mathcal{S} \subseteq N \times N$ will contain the pairs of norms that are connected by substitutabilities. In particular, notice that if $(n_i,n_j) \in R_s$, then $(n_i,n_j) \in \mathcal{S}$. 
$R_s$ is an irreflexive, symmetric, and transitive relation. 

\end{enumerate}

Next, we put together norms and their relationships in a structure on which reasoning about norms will take place, the so-called \emph{norm net}. 

\begin{mydef}
A norm net  is a pair \textit{NN}$=\langle N, R\rangle$, where $N$ stands for a set of norms and $R= \{R_g, R_x,R_s\}$ contains generalisation, exclusivity and substitutability relationships over the norms in $N$. The relationships in $R$ are mutually exclusive, namely $R_g \cap R_x= \emptyset$, $R_g \cap R_s= \emptyset$, and $R_x \cap R_s= \emptyset$. 
\end{mydef}

As observed in \cite{morales2015liberal}, generalisation and substitutability are mutually exclusive relationships. Furthermore, here we also assume that the exclusivity relationship is mutually exclusive with the generalisation and substitutability relationships.

Given a norm net \textit{NN}$=\langle N, R\rangle$, we will refer to any subset of the norms in $N$ as a \emph{norm system}. The challenge for a decision maker is to select a norm system out of a norm net. In general, she will be interested in norm systems that: (i) do not contain conflicting norms; (ii) do not contain overlapping regulations; and (iii) do incorporate as many norms as possible.
In what follows, we shall characterise these types of norm systems.



First, if we consider that exclusivity relationships capture conflicts between norms, the following characterisation of \emph{conflict-free} norm systems naturally follows.

\begin{mydef}
Given a norm net \textit{NN}$=\langle N, R\rangle$, we say that a norm system $\Omega \subseteq N$ is conflict-free iff for each $n_i, n_j \in \Omega$, $(n_i,n_j)\notin R_x$.
\end{mydef}

Second, notice that both generalisation and substitutability relationships capture \emph{redundancy}. Indeed, selecting two substitutable norms in a norm system would involve including overlapping regulations. The same applies to two norms such that one generalises the other, namely the regulation of the more general one subsumes that of the more specific. Hence, we characterise \emph{non-redundant} norm systems as follows.

\begin{mydef}
Given a norm net \textit{NN}$=\langle N, R\rangle$, we say that a norm system $\Omega \subseteq N$ is non-redundant iff for each $n_i, n_j \in \Omega$: (i) $(n_i,n_j)\notin R_g$ and $n_j \notin A(n_i)$; and (ii) $(n_i,n_j)\notin \mathcal{S}$
\end{mydef}







From the concepts above, we are ready to characterise the type of norm systems a decision maker will be interested in.


\begin{mydef}
Given a norm net \textit{NN}$=\langle N, R\rangle$, we say that a norm system $\Omega \subseteq N$ is sound iff it is both conflict-free and non-redundant.
\end{mydef}









\begin{figure}[tb]
\fbox{\includegraphics[width=3.2in, trim={3.8cm 6cm 3.5cm 5cm},clip] 
{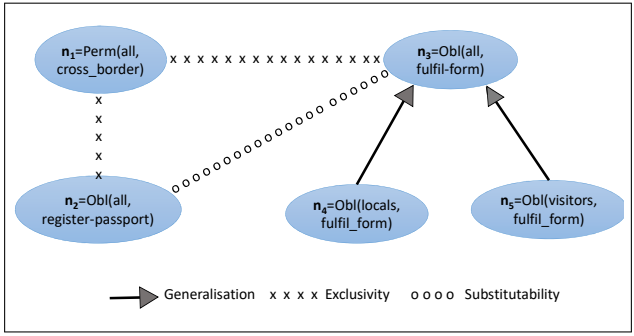}}
\centering
\caption{Norm Net example: rules of border control at an international airport.}
\label{fig:example}
\end{figure}
%
%
\begin{example}
Figure \ref{fig:example} illustrates an example of a Norm Net that includes some norms (rules) of border control at an international airport. Norms are depicted as circles labeled as $n_1, \ldots, n_5$ respectively. In particular, they are defined as follows:$\\
n_1: Permission (\mbox{\textit{all\_passengers, cross\_border}})\\
n_2: Obligation (\mbox{\textit{all\_passengers, register\_passport}})\\
n_3: Obligation (\mbox{\textit{all\_passengers, fulfil\_form}})\\
n_4: Obligation (\mbox{\textit{locals, fulfil\_form}})\\
n_5: Obligation (\mbox{\textit{visitors, fulfil\_form}})\\
$

Norm $n_1$ rules free movement of passengers, allowing all passengers to cross the border without any additional action. On the other hand, norm $n_2$ requires all passengers to register their passport, and there is still a third rule $n_3$ that requires them to fulfil a form asking for passport information such as passport number, holder's name or address. 

Regarding exclusivity relationships, first and second norms are exclusive $((n_1, n_2)\in R_x)$ because it is not possible to ask passengers to perform an action (in this case, register their passport) and allow them to go ahead and simply cross the border. The same reasoning applies for first and third norms, and thus, $n_1$ and $n_3$ are also exclusive $((n_1, n_3)\in R_x)$. Figure \ref{fig:example} depicts such exclusivity relationships with an ``x dotted'' line. 

Additionally, there is a substitutability relationship between second and third norms $((n_2, n_3)\in R_x)$, since it is possible to "monitor" which passengers are actually crossing the border by registering their password or by asking them to fulfil a form. Passengers may even have to abide by both norms, since they could actually do both things despite of its redundancy. Figure \ref{fig:example} shows this substitutability relationship with an ``o dotted'' line. 

Having a closer look to norm $n_3$, we can see that asking all passengers to fulfill a form is a generalisation of two other norms: $n_4$, which requires local passengers to fulfill a form; and $n_5$, which requires foreign passengers (visitors) to fulfill a form. Formally, we have $((n_3, n_4)\in R_g)$ and $((n_3, n_5)\in R_g)$. Figure \ref{fig:example} draws this generalisation relationship with an arrow line. Since generalisation is an anti-symmetric relationship, the arrowhead points towards the general norm.   
\end{example}


   

\section{Finding a maximum norm system}
\label{sec:maxNormSystem}
%
The purpose of this section is to design the optimisation machinery required to help a regulation authority find a particular type of sound norm system. Since the regulation authority's purpose is to incorporate as many norms as possible out of those proposed in a norm net, informally we will aim at the norm system that \emph{represents} the largest number of norms in the norm net. With this goal in mind, we first start in section \ref{subsec:mnsp} by formally casting our problem as an optimisation problem, and by characterising its hardness. Thereafter, we show how to encode our problem as a linear program so that it can be solved with the aid of state-of-the-art linear programming solvers. Finally, in section \ref{subsec:computingRepPower} we discuss on several ways of computing the representativeness of a norm system. 

\subsection{The maximum norm system problem}
\label{subsec:mnsp}

We first focus on defining how to obtain the representation power of a normative system. This is based on the representation power of each of its norms. Since we can think of several ways of computing a norm's representation power (e.g. a norm can represent itself, or all the norms it generalises), as discussed in section \ref{subsec:computingRepPower}, for now we just consider that we will count on a linear function, the so-called representation power function, $r:N \rightarrow \mathbb{R}$ that yields such value. Besides linearity, the only condition that we will impose on $r$ is that $r(n_i) \leq r(n_j)$ for each $n_j \in \mathcal{A}(n_i)$. Hence, the representation power of a normative system $\Omega$ can be readily obtained by adding the representation power of its norms, namely $\rho(\Omega) = \sum_{n \in \Omega} r(n)$.   

Now we are ready to formalise our optimisation problem. 

\begin{myProbl} 
Given a norm net \textit{NN}$=\langle N, R\rangle$ and a representation power function r, the maximum norm system problem (MNSP) is that of finding a sound norm system $\Omega$ with the maximum representation power, namely such that there is no other norm system $\Omega'\subseteq N$ such that $\rho(\Omega') > \rho(\Omega)$.  
\label{prob:mnsp}
\end{myProbl}

\begin{mylemma}
The complexity of the maximum norm system problem is at least NP-Hard.
\end{mylemma}
\begin{myproof}
The proof goes trivially by reduction of the maximum independent set problem, which is known to be an NP-Hard optimisation problem \cite{karp1972reducibility}, to the maximum norm system problem. Consider that we want to find the maximum independent set of a graph $G = (V,E)$. Now say that each vertex in $V$ stands for a norm and each edge in $E$ stands for an exclusivity relationship in $R_x$. From this follows, that finding the maximum independent set of $G$ amounts to solving the maximum norm set problem on the norm net $\langle V,\{R_x\}\rangle$, where the representation power function is defined as $r(v) = 1$ for each $v \in V$. 
\end{myproof}

Next we show how to solve the MNSP by encoding the optimisation problem as a linear program. Thus, consider a norm net \textit{NN}$=\langle N, R\rangle$, and a set of binary decision variables $\{x_1, \ldots, x_{|N|}\} $, where each $x_i$ encodes the decision of whether norm $n_i$ is selected (taking value $1$) for a norm system or not (taking value $0$). Thus, 
solving the MNSP amounts to solving the following linear program:
\begin{equation}
max \sum_{i=1}^{|N|} r(n_i) \cdot x_i  
\label{eq:MAXCCprobl}
\end{equation}
subject to the constraints that capture the generalisation, exclusivity, and substitutatibility relationships in the norm net, which we specify as follows:

\begin{itemize}
\item A family of \textit{generalisation constraints} to avoid redundancy. 
Such constraints impose that: 

\begin{itemize}

\item A norm cannot be selected together with any of the norms that it directly generalises. Given a norm $n_i$, the norms generalised by $n_i$ is defined as $Children(n_i) = \{n_j|(n_i,n_j) \in R_g\}$. Then, formally the following constraints must hold:
\begin{equation}
x_i + x_j \leq 1 \quad n_j \in Children(n_i) \ \ 1 \leq i \leq |N| 
\label{eq:Gconstrain1b}
\end{equation}

\item All the children of a norm cannot be simultaneously selected. Formally:
\begin{equation}
\smashoperator{\sum_{n_j \in Children(n_i)}} x_j < |Children(n_i)| \quad  1 \leq i \leq |N|
\label{eq:Gconstrain2b}
\end{equation}

\item A norm cannot be simultaneously selected together with any of its ancestors, namely:
\begin{equation}
x_i+x_k \leq 1 \quad n_k \in \mathcal{A}(n_i) \quad 1 \leq i \leq |N|
\label{eq:Gconstrain3b}
\end{equation}

\end{itemize}

\item \textit{Exclusivity constraints} preventing that two mutually exclusive (incompatible) norms are jointly selected to be part of a norm system. Thus, the following constraints must hold: 
\begin{equation}
x_i+x_j\leq 1 \quad \mbox{for each} \ (n_i,n_j)\in R_x
\label{eq:Xconstrainb}
\end{equation}
\item \textit{Substitutability constraints} avoiding that interchangeable norms are simultaneously selected. This amounts to enforcing that any pair of norms that are connected by substitutabilities cannot be simultaneously selected, namely:
\begin{equation}
x_i+x_j\leq 1 \quad \mbox{for each} \ (n_i,n_j)\in \mathcal{S}
\label{eq:Sconstrainb}
\end{equation}

Notice that this constraint ensures that two norms in $R_s$ are not jointly selected.
\end{itemize}

Furthermore, we must also consider the binary constraints corresponding to the norm decision variables, namely:
\begin{equation}
x_i \in \{0,1\} \quad 1 \leq i \leq |N|
\label{eq:binaryVariables}
\end{equation}

Notice also that the cost of encoding the MNSP as a linear program is  $\mathcal{O} (m \cdot |N| + \frac{|NN| (|NN| - 2)}{2})$, where $m = |R_g| + |R_x|$, and $\frac{|NN| (|NN| - 2)}{2})$ is the cost of encoding the substitutability constraints.


The specification above corresponds to a maximization problem whose constraints are all inequalities. Hence, it is in standard form and it can be solved with state-of-the-art linear program solvers such as CPLEX or Gurobi. 

\begin{example} Following our previous example in Figure \ref{fig:example}, we need five binary variables $(x_1, \ldots , x_5)$ to decide which norms to select to solve the MSNP.  If we consider $r(n_i) = 1$ for $1 \leq i \leq 5$, then there are four norm systems that maximise the objective function in equation \ref{eq:MAXCCprobl} and satisfy constraints \ref{eq:Gconstrain1b} to \ref{eq:binaryVariables}, namely: $\Omega_1=\{n_1, n_4\}$, $\Omega_2=\{n_1, n_5\}$, $\Omega_3=\{n_2, n_4\}$, and $\Omega_4=\{n_2, n_5\}$. These alternative norm systems are listed along the first column of Table \ref{table:probls}.  
\end{example}


Finally, it may be worth mentioning that prior to determining the maximum sound norm system, one may wonder if there always is a sound norm system. This leads us to pose next problem that, despite its simplicity, provides us with further insights into the problem under study.

 \begin{myProbl}
 Given a Norm Net \textit{NN}$=\langle N, R\rangle$ composed by set of norms $N$ and their relationships $R$, we aim at assessing if there is some sound (non-empty) $\Omega\subseteq N$.
 \label{SATprobl}
 \end{myProbl}

 Given the binary nature of the norm relationships, norm systems composed by any single norm from $N$ will always comply with constraints, and thus, there will always exist at least $|N|$ alternative sound norm systems.

\subsection{Computing representation power}
\label{subsec:computingRepPower}

As previously mentioned, we aim at computing the norm system that \emph{represents} the largest number of norms in the norm net, which may not be the same as taking the largest number of norms because we must consider generalisation relations. Hence, we introduce alternative definitions of the representation power function $r$ to compute the representation power of norms. 

In this paper we propose two different variations of the representation power function. First, the \emph{inclusion power} function considers as representation power the number of norms directly generalised or indirectly generalised (by being an ancestor) of a norm. Thus, it provides a local measure of representation power. Second, the \emph{generalisation power} function measures the representation power of a norm based on its position in a generalisation hierarchy: the deeper a norm is in the generalisation hierarchy (the farther from the most general norm), the lower its representation power. The next subsections detail both functions.
%
%

%
%
%
\subsubsection{Inclusion power}
The inclusion power representation function is a function $r_I: N \rightarrow \mathbb{N}$ that maps each norm to the number of norms it generalises. Given a norm, its generalisation power is obtained by combining the norms that it directly generalises together with the norms that its children generalise. If a norm does not generalise any other norm, we will set its inclusion power to $1$. This leads to the following recursive definition of inclusion power:

$$r_I(n_i) =
\left\{
	\begin{array}{ll}
		1  & \mbox{if } \not\exists r_j \in N, (r_i,r_j) \in R_g \\
		1 + \ \ \ \displaystyle \smashoperator{\sum_{n_j \in Children(n_i)}} r_I(n_j)& \mbox{otherwise}
	\end{array}
\right.
$$

In our example in figure \ref{fig:example}, the inclusion power of $n_3$ is 3, whereas it is $1$ for the rest of norms because none generalises other norms.

\subsubsection{Generalisation power}

The generalisation power representation function is a function $r_g: N \rightarrow (0,1]$ that maps each norm to its \emph{generalisation power}. Informally, the less ancestors a norm has, the higher its generalisation power. Thus, a norm that has no ancestors takes on the maximum generalisation power (1). 
The generalisation power lowers as norms are deeper in the generalisation hierarchy (i.e., are more and more specific, and hence have more ancestors). 
If the generalisation power of norm $n_i$ is larger than that of norm $n_j$, we will prefer to include $n_i$ in our norm system. This is because their generalisation powers tell us that $n_i$ is further up in the generalisation hierarchy than $n_j$, and hence closer to more general norms.

 
Algoritm \ref{alg:g} describes how generalisation power is computed for a Norm Net \textit{NN}$=\langle N,R\rangle$. Initially, lines \ref{alg-l:ini-g}-\ref{alg-l:ini-S} create two initial data structures: g[], a unidimensional array storing the representation powers for the norms in $N$; and $\mathcal{P}$, a set of the norms for which the generalisation power needs to be computed (initially $N$). Next, line \ref{alg-l:G} computes, by invoking function \textit{getNonGeneralised}(), the subset $G\subseteq N$ of norms not being generalised by any other norm (i.e.,
 $G \gets \{n_i  \in \mathcal{S} \; | \; \nexists n_j \; s.t. \; (n_j, n_i) \in R_g\}$).
Lines \ref{alg-l:forG1}-\ref{alg-l:forG2} assign a value of 1 for all the norms in $G$ and remove them from $\mathcal{P}$. The rest of the algorithm is devoted to repeat, while there are still norms in $\mathcal{S}$, the computation of the generalisation powers for subsequent levels in the generalisation hierarchy. Method \textit{SiblingGroups}($\mathcal{P}$, \textit{NN}) in line \ref{alg-l:GenSiblGroups} selects those norms that are generalised by norms that do not belong to $\mathcal{P}$ (i.e., $n_i$ s.t. $(n_j, n_i)\in R_g$ and $n_j\in N\setminus \mathcal{P}$) and groups them by siblings so that it returns \textit{SibList}[], a list of sets of siblings  $\mathcal{P}_s=\{n_{s1}, \ldots, n_{sk}\}$ s.t. $(n_j, n_{s1})\in R_g, \ldots,$  $(n_j, n_{sk})\in R_g$ for each common parent $n_j$. Then, line \ref{alg-l:gi} assigns, for each sibling norm $n_i \in \mathcal{P}_s$, a generalisation value $g[i] = \frac{g[j]}{|\mathcal{P}_s|}$, which corresponds to the generalisation power of its parent $n_j$\footnote{Notice that by construction, all norms in $N \setminus \mathcal{P}$ do have a computed generalisation power.} over the number of siblings. Once it is computed for all norms in $\mathcal{P}_s$, the entire subset is subtracted from $\mathcal{P}$ (see line \ref{alg-l:subsS}). Finally, the algorithm returns the g[] array that is used for the implementation of function $g_i(n_i)=$g[$i$]. Thus for example, if we have a norm $n_g$ generalising two other norms $n_{s1}$ and $n_{s2}$, and $n_{s1}$ in turn generalises two additional norms $n_{s11}$ and $n_{s12}$, we would have the following generalisation powers: $r_g(n_1)=1$; $r_g(n_{s1})=r_g(n_{s2})=1/2$; and $r_g(n_{s11})=n_g(n_{s12})=1/4$. 

\textcolor{black}{Notice that this algorithm assigns a value to all norms in $N$  and traverses in subsequent generalisation levels all the norms in a generalisation tree hierarchy, and thus, its complexity is $\mathcal{O}(l\cdot|R_g|+|N|)$ being $l$ the maximum generalisation level (i.e., the depth of the deeper generalisation hierarchy). }

\begin{algorithm}
  \caption{Generalisation power computation}\label{alg:g}
  \begin{algorithmic}[1]
    \Procedure{GeneralisationComp}{\textit{NN}}
      \State g[1..$|N|$] \label{alg-l:ini-g} \Comment{generalisation power array} 
      \State $\mathcal{P} \gets N$  \label{alg-l:ini-S} \Comment{set of pending norms}
      \State $G \gets$ \textit{getNonGeneralised}($\mathcal{P}$, \textit{NN}) 
      \label{alg-l:G}\Comment{general norms}
     \For{each norm $n_i \in G$}\label{alg-l:forG1}
      	\State g[$i$] $\gets 1$
        \State $\mathcal{P} \gets \mathcal{P} \setminus \{ n_i\}$ 
      \EndFor\label{alg-l:forG2}
      \While{$\mathcal{P}\neq\emptyset$}
        \State \textit{SibList}[] $\gets$ \textit{getSiblingGroups}($\mathcal{P}$, \textit{NN}) \label{alg-l:GenSiblGroups}
        \For{each $\mathcal{P}_s$ in \textit{SibList}[]}  
           \For{each sibling $n_i\in\mathcal{P}_s$}
              \State g[$i$] $\gets$ g[$j$]/$|\mathcal{P}_s|$ \label{alg-l:gi}
           \EndFor
           \State $\mathcal{P} \gets \mathcal{P} \setminus \mathcal{P}_s$ \label{alg-l:subsS}
        \EndFor
      \EndWhile\label{alg-l:GenDegCompWhile}      
      \State \textbf{return} g[1..$|N|$]\Comment{all values being defined}
    \EndProcedure
  \end{algorithmic}
\end{algorithm}

\begin{example} Following our example in Figure \ref{fig:example}, we can compare the two alternative representation power functions described above. On the one hand, if we employ inclusion power as representation function, the maximum norm system is $\Omega= \{n_3\}$. Naturally, $\Omega= \{n_3\}$ (see second column in Table \ref{table:probls}) has been preferred over the rest of sound norm systems, since  its representation power is 3, whereas the representation power for norm systems with two norms (e.g., $\Omega = \{n_1, n_4\}$) is 2.  
On the other hand, if we consider generalisation power instead, there are four equally-valued maximum norm systems, namely $\Omega_1=\{n_1, n_4\}$, $\Omega_2=\{n_1, n_5\}$, $\Omega_3=\{n_2, n_4\}$, and $\Omega_4=\{n_2, n_5\}$, as shown along the third column of Table \ref{table:probls}.
\end{example}

Overall, using alternative representation power functions results in different maximum norm systems. The one to use is a decision left to the decision maker. In any event, representing as many norms as possible may not always be the only criterion to enact a norm system. This is the case if the decision maker considers further criteria --such as, for instance, the costs associated to norms. Next section tackles this issue.

\section{Multi-objective optimisation for norm decision making}
\label{sec:OptProbl}
In addition to pursuing norm representation maximisation when choosing the norms to enact in a society, most often, regulation authorities cannot ignore the fact hat norm deployment has associated costs. This section deals with the combination of both criteria --i.e., representation power and associated costs-- to decide the norm system to enact.

Norm costs may represent monetary expenses derived from regulatory processes --such as norm establishment or norm enforcement-- as well as non-monetary aspects --such as social implications or political correctness-- that can be somehow quantified. Given a norm system, here we will consider that the cost of a norm system can be obtained by adding the value of its norms, namely $cost(\Omega) = \sum_{n_i \in \Omega}c(n_i)$, where $c(n_i)$ stands for the cost of norm $n_i$.
Furthermore, we make the (reasonable) assumption that costs are bounded by a maximum budget $b$ (i.e., the price regulatory authorities are willing to pay) that is available to cover the expenses of imposing those norms in the resulting Norm System. Then, we can cast the decision problem faced by the decision maker as the following multi-objective optimisation problem.

\begin{myProbl}
Given a Norm Net \textit{NN}$=\langle N, R\rangle$, a representation power function $r$, and a fixed budget $b$, the maximum norm system problem with limited budget (MNSPLB) is the problem of finding a sound norm system $\Omega\subseteq N$ with maximum representation power and minimum cost limited by some non-negative budget $b$. 
\label{OPTproblem}
\end{myProbl}



In order to solve the MNSPLB, we will try to cast it as a linear program, likewise we did for the MNSP in section \ref{subsec:mnsp}. Notice though that solving the MNSPLB amounts to solving the following optimisation problem:
\begin{equation}
\begin{multlined}
\; \; \; \; max \sum_{i=1}^{|N|} r(n_i) \cdot x_i\\
min \sum_{i=1}^{|N|} c(n_i) \cdot x_i \; \; \;\; \; \; \; \;
\end{multlined}
\label{eq:multiObjectiveFunction}
\end{equation}
subject to constraints \ref{eq:Gconstrain1b} to \ref{eq:binaryVariables}, namely the constraints that guarantee the soundness of the norm system, and a further constraint: 
\begin{equation}
\smashoperator{\sum_{i=1}^{|N|}} c(n_i) \cdot x_i \leq b,
\label{eq:limitedBudget}
\end{equation}
where $b \geq 0$, to ensure that the cost of the norm system does to go beyond the limited budget.
The multi-objetive optimisation problem represented by expression \ref{eq:multiObjectiveFunction} above can be formulated as a single-objective problem by aggregating the two objectives by means of scalarisation \cite{hwang2012multiple}. This is achieved by: (i) \emph{normalising} the representation and cost values; and (ii) \emph{prioritising} representation and cost values by means of weight values. On the one hand, we can readily normalise representation values by considering a maximum representation power. We can safely set this value to be $\mathcal{R}_{max} = \sum_{n_j \in G_N}r(n_j)$, where recall that $G_N$ stands for the set of norms that are not directly generalised by any other norm. On the other hand, cost values can be normalied by means of the maximum budget. As to how to prioritise representation costs and values, we can simply employ a weight $w_r$ to weigh the importance of maximising norm representation power, and another weight $w_c$ to weigh the importance of minimising norm costs.  

Putting all this together, the MNSPLB can be cast as a single-objective optimisation problem that can be solved by the following linear program:
\begin{equation}
max \; \Big[\, \frac{w_r}{\mathcal{R}_{max}} \cdot \sum_{i=1}^{|N|} x_i \cdot r(n_i) + w_c \cdot (y - \frac{1}{b}\sum_{i=1}^{|N|} x_i \cdot c(n_i) \Big]  
\label{eq:OPTproblem}
\end{equation}
subject to the constraints in equations \ref{eq:Gconstrain1b}-\ref{eq:binaryVariables}, the budget constraint in equation \ref{eq:limitedBudget}, plus the following additional constraints:

\begin{itemize}
\item The auxiliary indicator variable $y$ must be assigned as follows:
$$y =
\left\{
	\begin{array}{ll}
		1  & \mbox{if } \smashoperator{\sum_{i=1}^{|N|}} x_i > 0 \\
		0 & \mbox{Otherwise}
	\end{array}
\right.
$$
Notice that the assignment above captures the satisfaction of the following non-linear constraint: $\sum_{i=1}^{|N|} x_i > 0 \implies y=1$. Thus, the indicator variable will take on value $1$ the decision variables indicate that at least one norm is selected as part of the norm system, and value $0$ when the selected norm system is empty (namely $x_i = 0$ for all $1 \leq i \leq |N|$). In other words, the indicator variable $y$ is binary, and hence must satisfy that:
\begin{equation}
y \in \{0,1\}
\label{eq:yBinary}
\end{equation}

Thus, the indicator variable guarantees that nothing is added to the objective function if no norm is chosen. Furthermore, this indicator variable allows us to turn the minimisation problem in equation \ref{eq:multiObjectiveFunction} into a maximisation problem, since finding the norm system with minimum (normalised) cost ($\frac{1}{b}\sum_{i=1}^{|N|} x_i \cdot c(n_i)$) amounts to maximising expression $y - \frac{1}{b}\sum_{i=1}^{|N|} x_i \cdot c(n_i)$. 

However, notice that the addition of indicator variable $y$ requires the satisfaction of a non-linear constraint (i.e., an implication). Hence, in order to keep the problem linear, we will linearise such constraint as follows:
\begin{equation}
y \leq \sum_{i=1}^{|N|} x_i \leq  M \cdot y
\label{eq:Yconstrain}
\end{equation}
where $M$ is a very large number. \footnote{In our problem $M$ can be 
defined to be strictly larger than $|N|$.} 
\item The weights to measure the importance of maximising representation power and minimising cost must satisfy:
\begin{equation}
w_r + w_c = 1 \qquad w_r, w_c \in [0,1]    
\label{eq:Wconstrain}
\end{equation}
\end{itemize}

\textcolor{black}{The next example helps us illustrate how different elements in our multi-objective decision function influence the norms that are finally chosen to establish in a society. Specifically, we focus on the influence of: using our different representation power functions; changing norm costs; and variations in the total budget limit.} 

\begin{example} In our example, we could initially assume a maximum budget of 5 and consider the following costs: $n_1$ has no associated cost because it requires no additional actions; $n_2$ has an associated cost of 2 since it requires passengers to interact with passport registration machines and a few staff members; the cost of $n_3$ is 5 due to the fact that it requires form fulfilling, gathering, and post-processing; and $n_4$ and $n_5$, which are more specific than $n_3$, just cost 2. Hence: $b{=}5$, $c_1{=}0$, $c_2{=}c_4{=}c_5{=}2$, $c_3{=}5$. Moreover, in this example we will equally value the importance of norm representation and deployment cost, and thus, we set $w_r=w_c=0.5$, and, finally, the maximum representation power is computed as $\mathcal{R}_{max} = 5$ when considering $r_I$ whereas $\mathcal{R}_{max} = 3$ if $r_g$ is used instead. 

Now, considering $r_g$ --the generalization power function-- in equation \ref{eq:OPTproblem}, with restrictions from equations \ref{eq:Gconstrain1b}-\ref{eq:binaryVariables}, \ref{eq:limitedBudget}, and \ref{eq:yBinary}-\ref{eq:Wconstrain}, the linear program solver returns $\Omega= \{n_1\}$ as the optimally sound norm system (see first $\Omega$ in fifth column in Table \ref{table:probls}). Nevertheless, if cost of norm $n_1$ is increased up to $c_1=6$, since not checking passports at the border may cause political problems with the neighbouring countries, the solver will then choose $\Omega=\{n_2\}$ to be the optimally sound norm system (see second $\Omega$ in fifth column in Table \ref{table:probls}).

Alternatively, if we consider the inclusion power function $r_I$ instead, but keep initial costs assumptions (i.e.,$c_1=0$) $\Omega= \{n_1\}$ remains as the optimally sound norm system (as shown in first $\Omega$ in fourth column in Table \ref{table:probls}). In fact, this result does not change if we decrease the maximum budget to $b=4$. However, if we further increase it to $b=10$, this optimally sound norm system is enlarged to include one of the specific norms so that $\Omega=\{n_1, n_4\}$ or $\Omega=\{n_1, n_5\}$ in second $\Omega$ in Table \ref{table:probls}'s fourth column. Furthermore, the solver also computes different results for $c_1=6$ and maximum budgets: $b=4$ produces three possible optimally sound norm systems: $\Omega= \{n_2\}$, $\Omega=\{n_4\}$, and $\Omega= \{n_5\}$ (see third $\Omega$ in fourth column); $b=5$ implies two possible optimally sound norm systems $\Omega=\{n_4\}$ and $\Omega=\{n_5\}$; and finally, $b=10$ results in $\Omega=\{n_3\}$, since the inclusion power of $n_3$ can compensate its high cost ($c_3=5$) when normalised by 10.              
\end{example}
Previous example has briefly illustrated how several factors in our multi-objective decision function influence the computed result. Nevertheless, as previously mentioned, we assume the regulatory authorities' particular needs will naturally determine these factors. It is important to have in mind that our aim here is limited to model alternative problems, and thus, we summarise them in subsequent columns in Table \ref{table:probls}. Specifically, first raw describes in plain (abbreviated) language the criteria that has been applied for the formalisation of each problem. Then, second row references the equation that describes the objective function. Third row specifies the actual representation power function in use. Fourth row lists the constraints to which the objective function is subject to. And finally, last row compiles the different norm systems that have been computed for our running example.

\begin{table}
\caption{Proposed problems formulation and results for the example in Figure \ref{fig:example}.}
\label{table:probls}
\begin{tabular}{|l||c|c|c|c|}
\hline
 & \textit{MNSP}& \textit{MNSP}& \textit{MNSPLB}& \textit{MNSPLB}\\
 \hline
 & Max. & Max. & \parbox[t]{1.5cm}{Multi-objt.} & \parbox[t]{1.5cm}{Multi-objt.}\\
 \hline
  \parbox[t]{1.2cm}{Criteria} &
  \parbox[t]{0.5cm}{max.\\incl.} &
  \parbox[t]{0.5cm}{max.\\ gen.} &
  \parbox[t]{1.5cm}{max. incl.\\ min. cost} &
  \parbox[t]{1.5cm}{max. gen.\\ min. cost}\\
 \hline
 \parbox[t]{1.2cm}{Objective function} &
 \ref{eq:MAXCCprobl} &
 \ref{eq:MAXCCprobl} &
 \ref{eq:OPTproblem} &
 \ref{eq:OPTproblem}\\
 \hline
 \parbox[t]{1.2cm}{Represent. power \; $r$} &  \parbox[t]{0.5cm}{inclus.\\$r_I$} &  \parbox[t]{1cm}{generl.\\$r_g$} &  \parbox[t]{1.5cm}{inclusion\\$r_I$} &  \parbox[t]{1.5cm}{generalisat.\\$r_g$} \\
 \hline
  \parbox[t]{1.2cm}{Constraints} & \ref{eq:Gconstrain1b}-\ref{eq:binaryVariables} & 
 \parbox[t]{1cm}{\ref{eq:Gconstrain1b}-\ref{eq:binaryVariables}} & 
 \parbox[t]{1.5cm}{\ref{eq:Gconstrain1b}-\ref{eq:binaryVariables},\ref{eq:limitedBudget},\ref{eq:yBinary}-\ref{eq:Wconstrain}} & 
 \parbox[t]{1.5cm}{\ref{eq:Gconstrain1b}-\ref{eq:binaryVariables},\ref{eq:limitedBudget},\ref{eq:yBinary}-\ref{eq:Wconstrain}}\\
\hline
 \multirow{7}{*}{\parbox[t]{1.2cm}{Example solutions}} &
 \multirow{7}{*}{\parbox[t]{0.8cm}{$\Omega=\{n_3\}$}} & 
 \multirow{7}{*}{\parbox[t]{1cm}{$\Omega=$\\
 $\{n_1,n_4\}$, $\{n_1,n_5\}$, $\{n_2,n_4\}$, $\{n_2,n_5\}$}} &
 \parbox[t]{1.5cm}{$\Omega=\{n_1\}$}  &
 \parbox[t]{1.5cm}{$\Omega=\{n_1\}$}\\\cline{4-5}
 &  &  & \parbox[t]{2cm}{$\Omega=\{n_1,n_4\},$} &$\Omega=\{n_2\}$\\ \cline{5-5}
 &  &  & \parbox[t]{1.5cm}{$\{n_1,n_5\}$} &\\ \cline{4-4}
 &  &  & \parbox[t]{1.5cm}{$\Omega=\{n_2\},$} & \\
 &  &  & \parbox[t]{2cm}{$\{n_4\},\{n_5\}$} & \\\cline{4-4}
 &  &  & \parbox[t]{2.2cm}{$\Omega=\{n_4\},\{n_5\}$} &\\\cline{4-4}
 &  &  & \parbox[t]{1.5cm}{$\Omega=\{n_3\}$} & \\
\hline
\end{tabular}
\end{table}
\section{Moral values in norm decision making}
\label{sec:ValProbl}

%

So far we have considered quantitative criteria that regulation authorities can take into account when choosing the norms to enact in a society. However, decision makers may also require to assess the moral values promoted by 
 the norms in a given norm network under analysis. Here we understand that a norm supports a given moral value when the norm is meant to accomplish a goal that is aligned with (or promotes) the moral value. In this case, we will assume that the society has moral/social preferences over values. Thus, bringing in values into the norm decision making of regulatory authorities can be regarded as a qualitative criterion.

Based on the moral/social preferences over values, to reason about norm systems we must be able to compare them in terms of the values that they support. The principle that we adhere to is: \emph{the more preferred the values supported by a norm system, the more preferred that norm system}. Thus, ideally, the decision maker would like to opt for the norm system that supports the most preferred values out of all the sound norm systems. Nonetheless, we cannot forget that further criteria such as representation power and cost, as dicussed in sections \ref{sec:maxNormSystem} and \ref{sec:OptProbl}, must be part of the decision making. This section is devoted to extend the multi-objective decision-making problem introduced in section \ref{sec:OptProbl} to account for the moral \emph{values} supported by norms.

Our first goal is to be able to quantitatively reason about norm systems based on the qualitative preferences over the moral values that they support. First, in order to connect moral values and norms, we adapt some value-related definitions by Bench-Capon et al.\cite{Bench-CaponA09}. Thus, we shall consider $V$ as a non-empty set of moral values in a society. We will assume that values are not connected objects, namely there are no relationships between them, like e.g. complementarity. Moreover, we will assume that there is a total order (no ties) $\succ$ over the moral values in $V$ that reflects the moral/social preferences over them. Without loss of generality, we can assume that $v_1 \succ v_2 \succ \ldots \succ v_{|V|}$. Now, we can obtain the value(s) supported by each norm by means of a function $val:\mathcal{N} \rightarrow 2^V \setminus \emptyset$. Thus, $val(n_i)$ stands for the set of values promoted by norm $n_i$.

Now we introduce a utility function that will allow us to capture the total ordering over values. The utility of a value can be calculated as:
\begin{equation}
u(v_i)=  1 + \sum_{k = i+1}^{|V|} u(v_k)
\label{eq:utilityOfValues}
\end{equation}

where $1 \leq i \leq |V|$. From this definition, it is clear that $u(v_i) > u(v_j) \Leftrightarrow v_i \succ v_j$
From this, we can readily calculate the \emph{value support} of a norm $n_i$ by adding the utility of the values supported by the norm as follows: 
\begin{equation}
u_n(n_i) = \sum_{v \in val(n_i)} u(v)
\end{equation}

And from this, we can compute the \emph{value support} for a given norm system $\Omega \subseteq N$ by adding the utility of the values supported by each one of its norms as: 

\begin{equation}
u_N(\Omega) = \sum_{n \in \Omega} u_n(n)
\end{equation}

Notice that utility function $u_N$ allows us to lift the preferences defined as a linear order over single moral/social values to a preference relation over bundles of norms. Thus, we will say that $\Omega \succ \Omega' \Leftrightarrow u_N(\Omega) > u_N(\Omega')$. Interestingly, the lifting of preferences provided by the $u_N$ utility function satisfies two interesting properties: (i) \emph{responsiveness} \cite{Barbera2004}; and (ii) \emph{monotonicity}. Informally, responsiveness (also called pairwise-dominance), states that if in a norm system $\{n_2,n_3\}$, $n_3$ is replaced by a \emph{better} (supporting more preferred values) norm, e.g. $n_1$, then  $\{n_2,n_1\}$ makes a better norm system. Monotonicity states that if $\Omega \supset \Omega'$, then $\Omega \succ \Omega'$. These observations are fomally captured in the folllowing lemma.

\begin{mylemma}
The utility function $u_N$ guarantees responsiveness and monotonicity. 
\end{mylemma}
\begin{myproof}
To prove responsiveness it suffices to show that given a norm system $\Omega$ such that $n_i \in \Omega$, $n_j \not \in \Omega$, and $n_j \succ n_i$, then $\Omega \setminus \{n_i\} \cup \{n_j\} \succ \Omega$. Let us note $\Omega_{-i} = \Omega \setminus \{n_i\}$. Since $u_N(\Omega) = u_N(\Omega_{-i}) + u_n(n_i) < u_N(\Omega_{-i}) + u_n(n_j) = u_N(\Omega_{-i} \cup \{n_j\})$, then $\Omega \setminus \{n_i\} \cup \{n_j\} \succ \Omega$ holds.
As for monotonicity, this immediately follows from the definition of utility of a value in \ref{eq:utilityOfValues}. Given two norm systems such that $\Omega \supset \Omega'$, it is clear that $u_N(\Omega) > u_N(\Omega')$ since the value support for each norm in $\Omega \setminus \Omega'$ is greater or equal than $1$, and hence $\Omega \succ \Omega'$. 
\end{myproof}

At this point, we can quantitatively compare norm systems based on the values that they support. Hence, we are ready to define a new multi-objective optimisation problem involving values as an extension of problem \ref{OPTproblem}.

\begin{myProbl}
Given a Norm Net \textit{NN}$=\langle N, R\rangle$, a representation power function $r$, a fixed budget $b$, a set of values $V$ and a linear order $\succ$ over its values, the value-based maximum norm system problem with limited budget (VMNSPLB) is the problem of finding a sound norm system $\Omega\subseteq N$ with maximum representation power, minimum cost limited by some non-negative budget $b$, and maximum value support. 
\label{v-OPTproblem}
\end{myProbl}

This problem can be encoded as a linear program by extending the one in section \ref{sec:OptProbl}. In order to embed the maximisation of the value support of normative systems in the objective function of expression \ref{eq:multiObjectiveFunction}, we require: (i) a normalisation constant for the values of $u_N$; and (ii) some prioritisation weight $w_v$ that measures the relative importance of maximising value support. Notice that we can safely normalise moral values by considering  $\mathcal{V}_{max} = \sum_{i=1}^{|N|} u_n(n_i)$ as the normalisation constant. Then, solving the VMNSPLB amounts to solving the following linear program, which combines the maximisation of representation power, the minimisation of cost, and the maximisation of value support:

\begin{equation}
\begin{multlined}
max \; \Big[\, \frac{w_r}{\mathcal{R}_{max}} \cdot \sum_{i=1}^{|N|} x_i \cdot r(n_i) + w_c \cdot (y - \frac{1}{b}\sum_{i=1}^{|N|} x_i \cdot c(n_i)) + \\ 
+\frac{w_v}{\mathcal{V}_{max}} \cdot \sum_{i=1}^{|N|} x_i \cdot u_n(n_i)\Big]\; \; \;\; \; \; \; \;  
\label{eq:V-OPTproblem}
\end{multlined}
\end{equation}
subject to constraints from equations \ref{eq:Gconstrain1b} to \ref{eq:binaryVariables}, \ref{eq:limitedBudget},  \ref{eq:yBinary}, and \ref{eq:Yconstrain},
 together with a reformulation of the weight constraints (in eq. \ref{eq:Wconstrain}) as:
\begin{equation}
w_r + w_c + w_v= 1 \quad w_r, w_c, w_v \in [0,1]  
\label{eq:WconstrainV}
\end{equation}

\begin{figure}[tb]
\fbox{\includegraphics[width=3.2in, trim={3.8cm 3.5cm 2.5cm 3cm},clip] 
{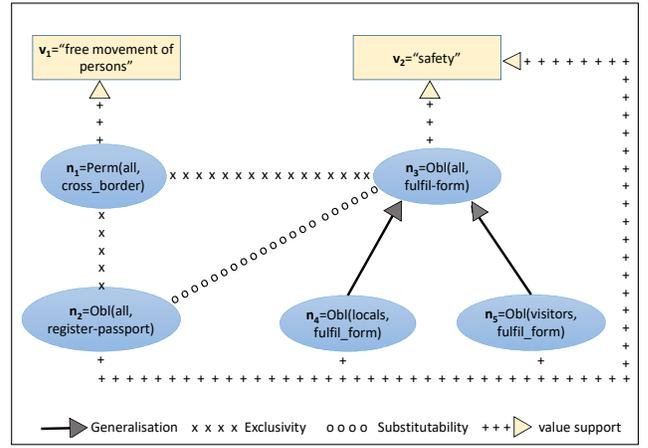}}
\centering
\caption{Example of rules of border control ($n_1, \ldots, n_5$) together with the values they support ($v_1, v_2$).}
\label{fig:Vexample}
\end{figure}

\begin{example} 
In our example, as Figure \ref{fig:Vexample} shows,  
we can consider that $n_1$ supports the  ``free movement of persons'' value ($v_1$), whereas $n_2, \ldots, n_5$ support the ``safety'' value ($v_2$). Let us consider that the society prefers ``free movement'' to ``safety'' (namely ``free movement'' $\succ$ ``safety''), then, $u(v_1) = 2$ and $u(v_2) = 1$ and that moral values are the only criterion to consider (i.e., $w_r=w_c=0$ and $w_v=1$). Therefore, our problem amounts to finding the sound norm system that has maximum value support. Then, if we encode the problem, a linear program solver results in 
two alternative solutions $\Omega=\{n_1, n_4\}$, $\Omega= \{n_1, n_5\}$.  
In other words, they constitute two different value-optimal sound norm systems.
\end{example} 





\section{Discussion: dealing with norms in force}
\label{sec:discussion}

So far we have considered that a norm net does contain all the norms and relationships a decision maker is to reason about. However, this many not be typically the case. Instead, she may already count on a collection of norms currently in force. If this is the case, the decision making process must consider such norms as part of the norm reasoning together with the norm net containing a new collection of candidate norms. Here we identify two main strategies to tackle this type of reasoning: (1) to consider that norms in force must be preserved; and (2) to consider that norms in force must not be necessarily preserved. The first strategy will have to consider that norms in force translate into \emph{hard constraints} in the norm decision making process, whereas the second strategy adds more flexibility to the decision maker. In what follows, we discuss how these strategies can be readily accommodated in our optimisation-based framework. 

Henceforth, we shall consider that $N_0$ stands for a set of norms in force and \textit{NN}$=\langle N, R\rangle$ stands for a norm net containing candidate norms to reason about together with their relationships. Whatever the strategy we adopt, we must build an \emph{extended norm net} that merges the norms in force with the norm net. This will result in a new norm net \textit{NN}$'=\langle N_0 \cup N, R \cup R_0\rangle$, where $R'$ stands for generalisation, exclusivity and substitutability relationships holding between the norms in force in $N_0$ and the norm candidates in the norm net, namely in $N$.

\begin{figure}[tb]
\fbox{\includegraphics[width=3.2in, trim={1.8cm 6cm 3.5cm 3.5cm},clip] 
{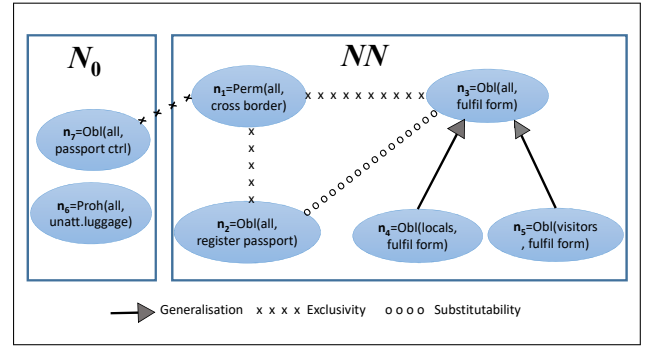}}
\centering
\caption{Example of $N_0$, a Norm Net \textit{NN'}, and an exclusivity relation between norms in $N_0$ and $N$.}
\label{fig:exampleP6}
\end{figure}

\begin{example} Figure \ref{fig:exampleP6} shows an extension of the example in figure \ref{fig:example}. The norms in force in $N_0$ are the following norms:
\\
$\\
n_6: Prohibition (\mbox{\textit{all\_passengers, unattend\_luggage}})\\
n_7: Obligation (\mbox{\textit{all\_passengers, passport\_control}})\\
$
\\
Norm $n_6$ prohibits all passengers to leave their luggage unattended and norm $n_7$ forces all passengers to go through passport control. Notice that there is a new exclusivity relation between $n_7$ and $n_1$, so that $(n_7,n_1)\in R_0$ will be added to the new norm net \textit{NN}'.
\end{example} 

Now we are ready to consider how to preserve the norms in force within our optimisation framework. This can be readily achieved by encoding the optimisation problem of choice for the extended norm net. However, imposing that the norms in force (i.e., in $N_0$) are always selected as part of the norm system requires to add further constraints to our optimisation problem. Thus, we add the following constraints:  
\begin{equation}
x_n = 1 \quad \mbox{for all} \; \; n \in N_0  
\label{eq:N0const}
\end{equation}

\vspace{-0.2cm}
\begin{example}Consider again our example in Figure \ref{fig:exampleP6}. Obviously, the norms to append to $N_0=\{n_6, n_7\}$ will depend on the prioritising criteria. However, norm $n_1$ in first $\Omega$ in Table \ref{table:probls}'s last column cannot be part of the resulting norm system because of its incompatibility with $n_7$. Hence, $n_2,n_3,n_4,n_5$ would constitute the norm candidates to be added to $N_0$. 
\end{example} 

Finally, consider that not all norms in force must be preserved. This adds more flexibility to the decision maker, who may discover a norm system that is actually \emph{better} (according to the decision criteria of choice) than the norm system resulting of the preservation of the norms in force. Notice that, in this case, we just need to encode our optimisation problem of choice for the extended norm net \textit{NN}' without adding any further constraints like \ref{eq:N0const}. 

\begin{example} 
Going back to our running example, if maximum budget is $b{=}5$, norms $n_6$ and $n_7$ have cost 1, $n_1$ has 0 cost, and we use $r_g$, then new norm net becomes  \textit{NN'}=$\langle\{n_1, n_6\},\emptyset\rangle$. As a consequence, $n_7$ would be replaced by $n_1$ in $N'$. Alternatively, in case we consider again $c_1{=}6$, no actual changes would take place in $N_0$ so that \textit{NN'}=$\langle \{n_6, n_7\},\emptyset\rangle$.
\end{example}

\section{Conclusions and future work}
\label{sec:concl}

The contribution of this paper is the modelling of a variety of norm decision-making problems faced by policy makers when choosing the norms to enact in a society. Our modeling assumes that 
a policy maker has knowledge about candidate norms to enact and the relationships between such norms. In particular, our work characterises generalisation, exclusivity, and substitutability as norm relationships. Furthermore, we show that the decision-making problems faced by policy makers can be cast as optimisation problems with multiple decision criteria (representation power, cost, and associated moral values). To the best of our knowledge, using moral values as a norm selection criterion is a particularly relevant contribution to the Normative Multi-Agent Systems research community.  Furthermore, the complexity of the resulting optimisation problems makes their solving non trivial. And yet, we managed to encode them as linear programs so that they can be solved with the aid of state-of-the-art linear programming solvers.

However, our work opens many interesting paths to future research. First, it is worth to conduct an empirical evaluation investigating the empirical hardness of different norm decision scenarios depending on the density of norm relationships. Second, reasoning about norms and values could be taken a step further by considering that norms support positively or negatively values. Moreover, from a pragmatic perspective: we could deal with a partial order (instead of a total order) over values, perform automated discovery of norm relationships, and embed our decision-making solvers into a decision support tool for policy makers.




\bibliographystyle{abbrv}
\bibliography{NormSynthesis}

\end{document}